\documentclass[12pt]{article}
\usepackage{amssymb,graphicx}
\usepackage{epstopdf}
\usepackage{amsmath,amsfonts}
\usepackage{epsfig}

\def\e{{\rm e}}

\def\d{\partial}
\def\l{\left(}
\def\r{\right)}

\newcommand{\be}{\begin{equation}}
\newcommand{\ee}{\end{equation}}
\newcommand{\bea}{\begin{eqnarray}}
\newcommand{\eea}{\end{eqnarray}}
\newcommand{\bg}{\begin{gather}}
\newcommand{\eg}{\end{gather}}
\newcommand{\bseq}{\begin{subequations}}
\newcommand{\eseq}{\end{subequations}}

\begin{document}
%\begin{flushright}
%Prepint Number
%\end{flushright}
\begin{flushright}
INR-TH-2015-010
\end{flushright}

\vspace{10pt}
\begin{center}
  {\LARGE \bf Conformal Universe  \\[0.3cm] as false vacuum decay } \\
\vspace{20pt}
%\medskip
M.~Libanov$^{a,b}$ and V.~Rubakov$^{a,c}$ %, S.~Sibiryakov$^{a,d,e}$\\
\vspace{15pt}

$^a$\textit{
Institute for Nuclear Research of
         the Russian Academy of Sciences,\\  60th October Anniversary
  Prospect, 7a, 117312 Moscow, Russia}\\
\vspace{5pt}

$^b$\textit{Moscow Institute of Physics and Technology,\\
Institutskii per., 9, 141700, Dolgoprudny, Moscow Region, Russia
}

\vspace{5pt}

$^c$\textit{Department of Particle Physics and Cosmology,
Physics Faculty, Moscow State University\\ Vorobjevy Gory,
119991, Moscow, Russia}

    \end{center}
    \vspace{5pt}

\begin{abstract}

We point out that the (pseudo-)conformal Universe scenario may
be realized as decay of conformally invariant, metastable vacuum,
which proceeds via spontaneous nucleation and subsequent growth
of a bubble of a putative new phase. We study perturbations about
the bubble and show that their leading late-time properties
coincide with those inherent in the original models with homogeneously
rolling backgrounds. In particular, the perturbations of a spectator
dimension-zero field have flat power spectrum.

\end{abstract}
\section{Introduction}

Unlike  inflationary scenario, conformal (or pseudo-conformal)
mechanism~\cite{Rubakov:2009np,Creminelli:2010ba,Hinterbichler:2011qk,Hinterbichler:2012fr}
attributes
the obesrved approximate flatness of the power spectrum of
cosmological
scalar perturbations to conformal symmetry %of an underlying theory
(see Ref.~\cite{Antoniadis:1996dj} for related discussion).
So far, the starting point has been time-dependent and spatially
homogeneous background, in which space-time is effectively
Minkowskian, while
conformal symmetry $SO(4,2)$ of a CFT is broken down to de~Sitter $SO(4,1)$
by the expectation value of a scalar operator
${\cal O}$ of non-zero conformal weight $\Delta$,
\be
\langle {\cal O} \rangle = \frac{\mbox{const}}{(-t)^{\Delta}}\; ,
\label{sep15-14-1}
\ee
where $t<0$. One introduces also another scalar operator $\Theta$, whose
effective conformal weight {\it in this background} is equal to zero, and whose
non-derivative terms in the linearized equation for perturbations
are negligible.
At late times,
the perturbations of $\Theta$ automatically have flat power spectrum.
If conformal symmetry is not exact at the rolling stage \eqref{sep15-14-1},
the spectrum is slightly tilted~\cite{Osipov:2010ee}.
It is furthermore assumed that the rolling regime \eqref{sep15-14-1}
holds only until some finite time $t<0$, and later on (possibly, much later)
the Universe enters
the usual radiation domination epoch,
cf.~Ref.~\cite{Wang:2012bq}. The perturbations of $\Theta$ are converted into
the adiabatic scalar perturbations after the rolling stage, by, e.g., one of the
mechanisms suggested in the
inflationary context~\cite{Linde:1996gt,Dvali:2003em}.

It is worth emphasizing that the (pseudo-)conformal mechanism can at best
be viewed as one of the ingredients of cosmological scenarios alternative to
inflation. It can work, e.g., at an early contracting stage of the
bouncing Universe~\cite{Khoury2004} or at an early stage of slow expansion in
the Genesis scenario~\cite{Creminelli2006,Creminelli:2010ba,Creminelli:2012my}.
These ``details'', however, are largely irrelevant as long as the
properties of perturbations are concerned, see Ref.~\cite{subscenarios}
for the discussion of possible sub-classes of the conformal mechanism.

A peculiar feature of the (pseudo-)conformal mechanism is that the
perturbations of ${\cal O}$ acquire {\it red} power spectrum,
\[
{\cal P}_{\delta {\cal O}} \propto k^{-2} \; .
\]
Interaction of the field $\Theta$ with the  perturbations
of ${\cal O}$ yields
potentially observable effects, such as statistical
anisotropy~\cite{Libanov:2010nk,subscenarios,Creminelli:2012qr,Ramazanov:2012za}
and specific shapes of
non-Gaussianity~\cite{Libanov:2011bk,subscenarios,Creminelli:2012qr}.
Many of these properties
are direct consequences of the symmetry breaking pattern
 $SO(4,2) \to SO(4,1)$~\cite{Hinterbichler:2012mv,Creminelli:2012qr}.

Overall, the (pseudo-)conformal scenario attempts to address the question:
``What if our Universe started off or passed through a conformally
invariant state and evolved into a much less symmetric state we see today?''
In this context, the spatially homogeneous Ansatz \eqref{sep15-14-1}
is rather ad hoc. Indeed, it would be more natural to think of
the rolling background similar to
\eqref{sep15-14-1} as emerging due to
a dynamical phenomenon associated with the instability of a conformally
invariant phase.
We propose in this paper that such a phenomenon may be
the decay of a metastable
(``false'') conformally invariant vacuum. A prototype example here
is a semiclassical scalar field theory with negative quartic potential
$V=-\lambda \phi^4$. In this theory, the metastable, conformally
invariant vacuum $\phi = 0$ decays via the Fubini--Lipatov
bounce~\cite{Fubini:1976jm,Lipatov:1976ny} whose interpretation is
the spontaneous nucleation of a spherical
bubble, which subsequently expands and whose interior has
the scalar field rolling down towards $\phi \to \infty$.
Clearly, this field configuration is not spatially homogeneous.

A somewhat more sophisticated construction involves holography.
False vacuum decay in adS has been studied in some
detail~\cite{Maldacena:1998uz,Gorsky,Hertog:2004rz,Craps:2007ch,Barbon:2010gn,Harlow:2010az}, with the emphasis on the (im)possibility of the resolution of the
big crunch singularity. Our prospective is different:  we treat the
false vacuum decay in
adS$_5$ as describing the instability of a conformally invariant
vacuum of the boundary CFT. Again, the resulting field configuration
is not spatially homogeneous, unlike in the earlier holographic approaches
to the (pseudo-)conformal
Universe~\cite{Hinterbichler:2014tka,Libanov:2014nla}.

A question naturally arises as to whether the false vacuum decay picture
of the \\ (pseudo-)conformal Universe yields the same predictions for field
perturbations as the homogeneous rolling model
\eqref{sep15-14-1}: flat and red power spectra of perturbations of the
fields $\Theta$ and ${\cal O}$, respectively. It is this question that we
mainly focus on in this paper.
By explicitly considering the examples alluded to above,
we find that the perturbations of relevant wavelengths do have these
properties.
We will see that this feature is again dictated by
the symetry breaking pattern, in particular, by spontaneously broken
dilatation invariance.  Thus, the potentially observable features of the
(pseudo-)conformal Universe, studied in the context of the homogeneous model
\eqref{sep15-14-1}, hold also in the false vacuum decay
scenario.

The paper is organized as follows. In Section~\ref{sec:toy} we consider
a toy model of a scalar field with negative quartic potential in flat
space-time. To have conformal symmetry, we treat this model at
semiclassical level. After recalling the Fubini--Lipatov bounce solution
in Section~\ref{sec:toybubble}, we consider perturbations of the
scalar field itself in Section~\ref{subsec:radial} and of a spectator
dimension-zero field in Section~\ref{dim0}. We find that at late times,
these perturbations have red and flat spectra, respectively. We make a
few remarks in Section~\ref{sub:remarks}. A holographic model
is studied in Section~\ref{sec:holog}, where we work exclusively
in the probe scalar field approximation and hence neglect the
deviaton of metric from adS$_5$. We introduce the bounce configuration
 in adS$_5$ and discuss its properties in Section~\ref{sub:adSbounce}.
The two types of perturbations are considered in Sections~\ref{subsec:adSradial}
and \ref{subsec:adSphase}, respectively. We identify the 5d modes
that dominate at late times and show that  they
again have red and flat spectra, respectively.
We conclude in Section~\ref{sec:concl}.

\section{Toy model: semiclassical $(-\lambda \phi^4)$ theory}
\label{sec:toy}

\subsection{Nucleated bubble}
\label{sec:toybubble}

To begin with, let us consider a semiclassical scalar field theory in
4d Minkowski space, with the action  (mostly positive signature)
\be
S = \int~d^4 x~\left( - \frac{1}{2}\d_\mu \phi \d^\mu \phi +
\frac{\lambda}{4}\phi^4 \right) \; .
\label{dec17-14-5}
\ee
The conformally invariant vacuum $\phi = 0$ is perturbatively stable
(small local perturbations about this vacuum have positive gradient
energy
at quadratic level)
and  decays via the Lipatov--Fubini
bounce (instanton)~\cite{Fubini:1976jm,Lipatov:1976ny},
the following solution to the Euclidean
field equation:
\be
\phi_c = \sqrt{\frac{8}{\lambda}} \frac{\rho}{\rho^2 + x^2} \; ,
\label{dec11-14-1}
\ee
where $\rho$ is an arbitrary parameter, the instanton
size.
The action for this instanton equals $S_I = 8\pi^2/\lambda$.
Upon analytical continuation $x^0 \to ix^0$, the solution
has the same form as \eqref{dec11-14-1} but with Minkowskian $x^2$.
It describes a bubble that materializes at $x^0 = 0$ and expands
afterwards.
Inside the bubble, the field rolls down towards $\phi \to \infty$,
and (formally) reaches infinity at $x^2 = -\rho^2$. The
bubble nucleation and its subsequent expansion is a
prototype example of (pseudo-)conformal cosmological stage emerging in the
process of false vacuum decay. Similarly to the original scenario
\eqref{sep15-14-1}, we assume
that the rolling stage \eqref{dec11-14-1} terminates at a hypersurface
$x^2 = \mbox{const} > - \rho^2$.

A remark is in order. The vacuum decay rate in
our semiclassical toy model, as it stands, is UV divergent.
Indeed, if one insists on conformal invariance and,  for that matter, ignores
the renormalization group effects, one has for
the decay rate per unit time per unit volume
\be
\Gamma = \int \frac{d\rho}{\rho^5} \e^{-\frac{8\pi^2}{\lambda}} \; ,
\label{dec11-14-3}
\ee
where the integral diverges as $\rho \to 0$. This problem can be cured by
mild explicit breaking of conformal invariance (recall that
breaking of conformal invariance is in any case required for
obtaining the observed tilt of the scalar power spectrum). As an example,
the quartic coupling may depend on the scale in a way reminiscent
of the renormalization group evolution,
\be
\lambda^{-1} (\rho) = \left\{ \begin{aligned} &\lambda^{-1} \; , \;\;\;\;\;\;
&&\rho \gtrsim \mu^{-1}\\
&\lambda^{-1} + \beta \log(\rho \mu)  , \;\;\;\;\;\;
&&\rho \lesssim \mu^{-1} \; .
\end{aligned}
\right.
\label{dec11-14-2}
\ee
For
$\beta < -(2\pi^2)^{-1}$ the integral in eq.~\eqref{dec11-14-3} converges,
and the typical instanton size is $\rho \sim \mu^{-1}$.

\subsection{Radial perturbations}
\label{subsec:radial}

Let us now consider perturbations about the Minkowski
bubble solution,
$\phi = \phi_c + \delta \phi (x)$. We call them radial perturbations, cf.
Refs.~\cite{Rubakov:2009np,Libanov:2010nk}.
The quadratic action is
\[
S^{(2)}_{\delta \phi} = \int~d^4x~\sqrt{-g}~\frac{1}{2}~
\left( -g^{\mu \nu} \d_\mu \delta \phi\cdot \d_\nu \delta \phi +
\frac{24 \rho^2}{(\rho^2 + x^2)^2} (\delta \phi)^2 \right) \; ,
\]
where $g_{\mu \nu} = \eta_{\mu \nu}$ in Cartesian coordinates. We are
interested in short modes whose wavelengths are much smaller than $\rho$.
In the cosmological context this is justified by the fact that the
Universe filled with the rolling field \eqref{dec11-14-1} is homogeneous
on hypersurfaces $x^2 = \mbox{const}$, whose curvature radius is of order
$\rho$ towards the end of rolling stage, $x^2 \sim - \rho^2$, while the
scales of relevant perturbations are much shorter than the radius of
spatial curvature today, and hence at early times.

The short
modes start to
feel the background towards the end of rolling, when $\rho^2 + x^2 \ll \rho^2$,
i.e., at negative $x^2$. In this patch of Miknkowski space, the
convenient coordinates are $v$ and $\psi$ related to the
Cartesian radial coordinate and time by
%\begin{subequations*}
\begin{align*}
R \equiv \sqrt{{\bf x}^2} &= \rho_0 \e^{v} \sinh \psi
\\
x^0 &= \rho_0 \e^v \cosh \psi \; .
\end{align*}
%\end{subequations*}
We have chosen the coordinate transformation independent
of the instanton size $\rho$, so the parameter $\rho_0$ is an arbitrary
length scale. This will enable us to vary the instanton size without
touching the coordinate frame.
Note that $v\to - \infty$ corresponds to $x^2 \to 0$
and $v\to \log (\rho/\rho_0)$ corresponds to the (would-be)
end-of-roll hypersurface
$x^2 = -\rho^2$.
In these coordinates the Minkowski metric is
\be
ds^2 = \rho_0^2 \e^{2v} \left( - dv^2 + d\psi^2 + \sinh^2 \psi d\Omega_2^2
\right)
%\nonumber \\
 =
  \rho_0^2\e^{2v}(-
dv^2 +  \gamma_{ij} dX^i dX^j) \; ,
\label{sep24-14-3}
\ee
where $d\Omega_2^2$ is metric on unit 2-sphere and $ \gamma_{ij} $
is metric on unit 3-hyperboloid with coordinates $X^i$. Note that
$v$ is the time coordinate in the patch we consider.

We introduce new field variable via
\be
\sigma = \rho_0\e^v \delta \phi \; .
\label{sep24-14-2}
\ee
Then the action for perturbations becomes
\be
S^{(2)} = \int~dv~d^3X~\sqrt{\gamma}~\frac{1}{2}
~\left[ \left(\frac{\d \sigma}{\d v}\right)^2
+ \sigma^2 - \gamma^{ij} \frac{\d \sigma}{dX^i}
  \frac{\d \sigma}{dX^j}
+ 24 \frac{\rho^2\rho_0^2\e^{2v}}{(\rho^2- \rho_0^2\e^{2v})^2} \sigma^2 \right]
\ee
and  the linearized field equation reads
\be
\frac{\d^2 \sigma}{\d u^2} + (k^2 - 1) \sigma
- \frac{6}{\sinh^2 u} \sigma = 0 \; ,
\label{sep14-14-1a}
\ee
where $(-k^2)$ is an eigenvalue of the Laplacian on unit
3-hyperboloid and
\[
u = v - \log (\rho/\rho_0) \; .
\]
The coordinate $u$ runs from $u \to -\infty$, while the
end of roll  is at $u=0$.

In accord with the above discussion,
the short modes, $k \ll 1$, are in the WKB regime
until $u$ gets small,
$u \sim k^{-1} \ll 1$.
So, we can safely
take the small-$u$ asymptotics of eq.~\eqref{sep14-14-1a}:
\[
\frac{\d^2 \sigma}{\d u^2} + k^2 \sigma
- \frac{6}{u^2} \sigma = 0 \; .
\]
This equation is familiar in the context of
the (pseudo-)conformal scenario. As usual, its solution should tend  to
$\frac{\e^{-iku}}{\sqrt{2k}} B_{\bf k}$ at large negative $u$,
where $B_{\bf k}$ is an annihilation operator. This solution is
$\frac{\sqrt{\pi u}}{2} H^{(1)}_{5/2} (-ku)  B_{\bf k}$, and
at $k|u| \ll 1$ the field asymptotes to
\be
\sigma = -i \frac{3}{\sqrt{2} k^{5/2} u^2}  B_{\bf k} + \mbox{h.~c.}\; .
\label{dec17-14-1}
\ee
Assuming that the field $\sigma$ is in its
vacuum state at large negative
$u$, one finds that
the radial perturbations at $k|u| \ll 1$ have red power spectrum,
\be
{\cal P}_\sigma = \frac{9}{16\pi^2}\frac{1}{u^4 k^2} \; .
\label{jan9-15-5}
\ee
The perturbation \eqref{dec17-14-1} with its
time-dependence $\sigma \propto u^{-2}$ can be understood as
a coordinate-dependent rescaling of the bubble size. Indeed, consider
the bubble whose size $\rho$ slowly varies across the
3-hyperboloid. In coordinates $(v, {\bf X})$ this configuration is
\[
\phi = \sqrt{\frac{8}{\lambda}} \frac{\rho({\bf X})}{\rho^2 ({\bf X})-
\rho_0^2 \e^{2v}} \; ,
\]
where $\rho({\bf X}) = \rho + \delta \rho({\bf X})$. This configuration is
a perturbation about the background \eqref{dec11-14-1} with
\[
\delta \phi = -  \sqrt{\frac{8}{\lambda}}
\frac{2\rho}{(\rho^2 -
\rho_0^2 \e^{2v})^2} \cdot \delta \rho ({\bf X}) + \dots
= - \sqrt{\frac{2}{\lambda}} \frac{1}{\rho u^2}  \delta \rho ({\bf X}) + \dots
\; ,
\]
where dots stand for terms which are less singular as $u \to 0$.
Comparing this with \eqref{sep24-14-2} and \eqref{dec17-14-1} we see that
the field $\delta \rho$ is independent of time $u$ as $u \to 0$,
\[
\delta \rho =  - \sqrt{\frac{\lambda}{2}} u^2 \sigma
=  i \sqrt{\frac{\lambda}{2}} \frac{3}{\sqrt{2} k^{5/2}}  B_{\bf k}
 + \mbox{h.~c.}\; .
\]
It has red power spectrum
\[
{\cal P}_{\delta \rho} = \frac{9\lambda}{32 \pi^2} \frac{1}{k^2} \; .
\]
This observation helps understand the late-time behavior of the
perturbations, eq.~\eqref{dec17-14-1}.The
spatial gradients of $\delta \phi$ are
negligible in the late-time regime, and $\delta \phi$ is a solution
to the equation for homogeneous perturbation (in coordinates $(u, {\bf X})$).
Due to invariance under dilatations, one such solution is necessarily
$\d \phi_c / \d \rho$, while
another is less singular as $u \to 0$. Therefore, at
late times one has
$\sigma \propto \delta \phi \propto \d \phi_c / \d \rho \propto
u^{-2}$. The dependence on $k$ in eq.~\eqref{dec17-14-1} is then
restored on dimensional grounds. We conclude that the reason behind the
red power spectrum \eqref{jan9-15-5} is invariance under dilatations.

\subsection{Spectator dimension-zero field}
\label{dim0}

Let us now consider a spectator field $\Theta$
that has  zero effective conformal dimension in the background
\eqref{dec11-14-1}. While the existence of dimension-zero fields
in conformally invariant vacuum is problematic, such fields
may naturally exist when the background spontaneously breaks
conformal invariance. As an example, one can modify the model
\eqref{dec17-14-5} by considering complex field $\phi$. Then the
background
solution is still given by eq.~\eqref{dec11-14-1}, while the phase
$\Theta = \mbox{Arg}~ \phi$ automatically has dimension zero.

In any case, modulo overall
constant factor, the quadratic action for a spectator dimension-zero
field in the background
$\phi_c$ is
\[
S_\Theta = - \int~d^4 x~ \sqrt{-g} ~ \phi_c^2 (x) g^{\mu \nu}
\d_\mu \Theta \d_\nu \Theta \; ,
\]
where we assume that non-derivative terms for $\Theta$ are absent
(incidentally, this is the case for $\Theta = \mbox{Arg} ~\phi$).
From now on it is convenient to work with
the coordinate $u$, so that the metric
and background solution are
\begin{align*}
ds^2
& = \rho^2 \e^{2u}(-du^2 + \gamma_{ij}d X^i dX^j)
\\
\phi_c &=  \sqrt{\frac{8}{\lambda}} \frac{1}{\rho (1- \e^{2u})} \; .
\end{align*}
Upon introducing a field $\xi$ via
\[
\Theta = \sqrt{\frac{\lambda}{4}} ~\sinh u~\cdot \xi \; ,
\]
we obtain the action
\[
S_\xi = \int~du~d^3X~\sqrt{\gamma} ~\frac{1}{2}
\left[ \left(\frac{\d \xi}{\d u}\right)^2
+ \xi^2 - \gamma^{ij} \d_i \xi \d_j \xi + \frac{2}{\sinh^2 u} \xi^2 \right]
\; .
\]
Again considering the high momentum modes, $k \gg 1$, and hence late times,
$|u| \ll 1$, we arrive at the familiar equation
\[
 \frac{d^2 \xi}{d u^2} + k^2\xi - \frac{2}{u^2} \xi = 0 \; .
\]
Its properly normalized solution is
$\frac{\sqrt{\pi |u|}}{2} H^{(1)}_{3/2} (-ku)
A_{\bf k}$, where $A_{\bf k}$ is another annihilation operator.
At $k|u| \ll 1$ the field is
\[
\xi = -i \frac{1}{\sqrt{2} k^{3/2} u}  A_{\bf k} + \mbox{h.~c.}\; ,
%\label{dec17-14-7}
\]
so the field $\Theta$ is independent of time,
\[
\Theta = -i \sqrt{\frac{\lambda}{8}} \frac{1}{k^{3/2}}  A_{\bf k} + \mbox{h.~c.}
\; .
\]
It has flat power spectrum,
\be
{\cal P}_\Theta = \frac{\lambda}{16 \pi^2} \; .
\label{feb10-15-1}
\ee
If $\Theta$ is interpreted as the phase of the field $\phi$,
its independence of time at late times can be understood as a consequence
of the phase rotation symmetry $U(1)$ spontaneously broken by the
background \eqref{dec11-14-1}, cf. Refs.~\cite{Rubakov:2009np,Libanov:2010nk}.

\subsection{Remarks}
\label{sub:remarks}

To summarize, the late-time properties of the radial field $\delta \phi$
and spectator field $\Theta$, as given by eqs.~\eqref{jan9-15-5} and
\eqref{feb10-15-1}, are the same as in the homogeneous rolling background
\eqref{sep15-14-1}. In fact,
the example we have considered in this Section is quite trivial:
in the regime we have discussed it
reduces to the spatially homogeneous model with
$\phi_c \propto 1/(-t)$. Towards the end of rolling, when
$x^2 \to - \rho^2$, the solution \eqref{dec11-14-1} is approximately
\be
\phi_c = \sqrt{\frac{2}{\lambda}} \frac{1}{\rho - \tau} \; ,
\label{jan9-15-1}
\ee
where $\tau = \sqrt{-x^2}$. At that time, the variable $\tau=\rho_0 \e^v$
can be viewed as  time coordinate, and since $v \to \mbox{const}$,
the metric \eqref{sep24-14-3} is effectively static. Unlike in the
homogeneous rolling case, it has negative
 spatial curvature, but we considered short modes and  neglected
the spatial curvature anyway. So, the limit we have studied indeed
boils down to
the homogeneous model of Refs.~\cite{Rubakov:2009np,Hinterbichler:2011qk}.
This is further illustrated by the fact  that in this limit,
the symmetry under
rescaling of $\rho$, which is
the reason behind the red power spectrum of $\delta \phi$,
is equivalent to the symmetry under time shift (see
eq.~\eqref{jan9-15-1}), which is
responsible for the red spectrum of radial perturbations
in homogeneous
models~\cite{Rubakov:2009np,Creminelli:2010ba,Hinterbichler:2011qk}.
Thus, even though our starting point has been somewhat different,
physics of perturbations
is essentially the same as in the  homogeneous
rolling setup.

\section{Holographic model}
\label{sec:holog}

In this Section we consider a holographic model for the decay
of a conformally invariant false vacuum.
Like in refs.~\cite{Hinterbichler:2014tka,Libanov:2014nla},
we adopt a bottom-up approach, and instead of constructing
a concrete 4d CFT and its dual,
study a fairly generic 5d theory of a scalar field with action
\be
S = \int~\sqrt{-g} \l -\frac{1}{2} g^{AB} \d_A \phi \d_B \phi - V(\phi) \r
~ dz~d^dx \; ,
\label{may10-6}
\ee
living in adS$_5$ space with metric
(hereafter the adS$_5$ radius is set equal to 1)
\be
ds^2_5 = \frac{1}{z^2}\left(
\eta_{\mu \nu} dx^\mu dx^\nu + dz^2\right) \; .
\label{may10-2}
\ee
We will work in the probe scalar field approximation throughout,
so this metric is unperturbed.
The unstable conformally invariant vacuum is at $\phi=0$.
We would like this theory to correspond to a boundary CFT
without explicit breaking of conformal invariance. So, unlike in
Ref.~\cite{Distler:1998gb}
we assume that
the potential $V(\phi)$ has a {\it local} minimum at $\phi = 0$.
We will use the operator correspondence between the tree level theory
\eqref{may10-6} in adS$_5$ and large-$N$ CFT, as discussed in detail
in Ref.~\cite{Sundrum:2011ic}.
The field behavior near the adS boundary $z=0$ is
\be
\phi (z, x) = z^{\Delta_+} \phi_0 (x)
\label{dec10-14-1}
\ee
with
\be
\Delta_+ = \sqrt{m^2+4} +2 \; ,
\label{jan14-15-1}
\ee
where $m$ is the scalar field mass in the vacuum $\phi =0$, and
$\phi_0$ is related to
a CFT operator \cite{Balasubramanian:1998de,Klebanov:1999tb},
\be
{\cal O} = 2\sqrt{m^2 + 4} ~\phi_0 \; .
\label{dec10-14-3}
\ee
The property \eqref{dec10-14-1} implies that there is no explicit
deformation of the boundary CFT, in contrast to
Refs.~\cite{Hertog:2004rz,Craps:2007ch}.
We assume for definiteness (although this is not obligatory) that  the
potential $V(\phi)$ has a global minimum at $\phi = \phi_+ \neq 0$. Yet
another assumption, whose significance will become clear later, is that
the curvature of the potential at its maximum is large enough,
 \be
|V^{\prime \prime}(\phi_{max})| > 4 \; .
\label{dec10-14-2}
\ee
So, our model is essentially the same as in
Ref.~\cite{Libanov:2014nla}.

\subsection{Bounce}
\label{sub:adSbounce}

We begin with discussing an analog of the Fubini--Lipatov bounce,
a Euclidean solution describing the
tunneling process leading to the nucleation of a bubble in the
false vacuum. The Euclidean version of the theory is described by the
metric \eqref{may10-2} with $\eta_{\mu \nu} \to \delta_{\mu \nu}$,
or
\be
ds_5^2 = \frac{1}{z^2}(dz^2 + dR^2 + R^2 d\Omega_3^2 ) \; ,
\label{jan13-15-1}
\ee
where $d\Omega_3^2$ is the metric on unit 3-sphere.
For our purposes, however, a convenient form of the metric is
%\be
%\label{newmetr}
\be
ds^2_5=ds^2+\sinh^2 s(d\psi^2+\cos^2\psi ~d\Omega_3^2)\; .
\label{jan13-15-2}
\ee
The relation between coordinates in eqs.~\eqref{jan13-15-1} and
\eqref{jan13-15-2} is~\cite{Rubakov:1999ir},
\be
\label{jan13-15-3}
z=\frac{1}{\cosh s-\sin \psi\sinh s}~,~~~
R=\frac{\cos\psi\sinh s}{\cosh s-\sin \psi\sinh s}\;.
\ee
%\ee
In coordinates $(s, \psi)$,
slices of constant $s$ are 4-spheres; the unusual choice of the
coordinate $\psi$ is made for future convenience. An $SO(5)$-symmetric
bounce configuration depends on $s$ only,
$\phi_c = \phi_c (s)$, and the Euclidean action reads
\begin{equation*}
%\label{newS}
S=\frac{8\pi^2}{3}\int ds \sinh^4 s ~\left[
\frac{1}{2}\left(   \frac{d\phi_c}{ds}\right)^2+V(\phi_c) \right] \; ,
\end{equation*}
leading to the equation of motion
\begin{equation}
\label{neweqm}
\frac{d^2\phi_c}{ds^2}+4\frac{\cosh s}{\sinh s}\frac{d\phi_c}{ds}-V'(\phi_c)
=0\; .
\end{equation}
Interestingly, this equation coincides with the equation
for spatially flat domain wall  (in that case $\phi_c = \phi_c (\tilde{s})$,
$\cosh\tilde{s} = - t/z$, $t<0$), studied in
Ref.~\cite{Libanov:2014nla}.
We repeat here the argument~\cite{Libanov:2014nla}
showing that for a class of potentials, eq.~\eqref{neweqm} admits
a non-singular solution with $\phi_c \to 0$ as $s\to \infty$.
This is precisely the bounce solution we are after.

\begin{figure}[tb!]
\begin{center}
\includegraphics[width=0.6\textwidth,angle=0]{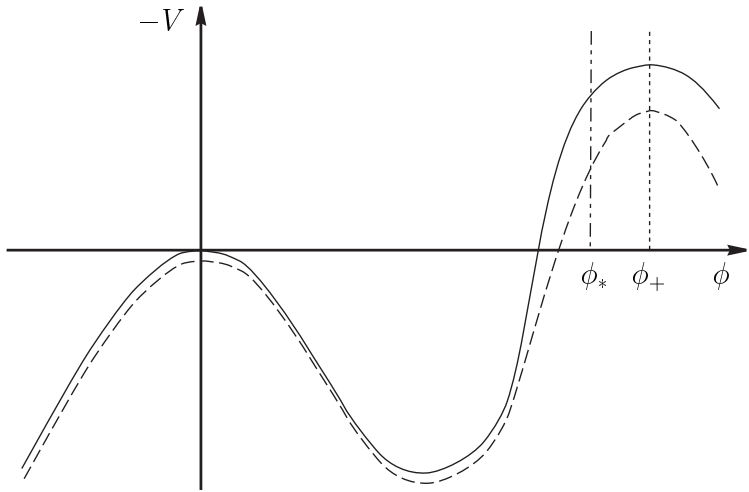}
\end{center}
\caption{Inverted potential $[-V(\phi)]$ (solid line)
and inverted auxiliary potential
$[-V_0 (\phi)]$ (dashed line).
\label{potential}
 }
 \end{figure}

Near $s=0$, the metric \eqref{jan13-15-2} is 5d flat, and $s$ serves as
the radial coordinate. Hence, the solution should obey
$d\phi_c (0)/ds = 0$. Its behavior at small $s$ is determined by the value
of $\phi_c (0) \equiv \phi_*$:
\[
\phi_c (s) = \phi_* + \frac{1}{10} V'(\phi_*)\cdot s^2 + O(s^4) \; .
\]
Now, eq.~\eqref{neweqm} corresponds to motion of a ``particle'' in the
inverted potential $-V$ with ``time''-dependent friction,
from $s=0$  to $s \to \infty$. As outlined above, we
consider the potentials $V(\phi)$ of
the form shown in Fig.~\ref{potential} (solid line),
and the ``initial''
values of $\phi_*$
to the right of the maximum of $V$ (minimum of $-V$). We would
like the solution to overcome the maximum of $V$ at some ``time''
$s_m$. Near the maximum $\phi_{max}$, the solution is a linear combination
 $\phi-\phi_{max} =c_1 \e^{-\gamma_1 s}+ c_2 \e^{-\gamma_2 s}$, where
 $\gamma_{1,2} = 2\coth s_m \pm \sqrt{4 \coth^2 s_m
+ V^{\prime \prime}(\phi_{max})}$.
Hence, the solution can overcome the maximum only for
 $V^{\prime \prime}(\phi_{max})<- 4$, otherwise
both $\gamma_1$ and $\gamma_2$ are real and positive, and the solution
gets stuck at
$\phi_{max}$. This is the reason for our assumption \eqref{dec10-14-2}.

Let $V_0 (\phi)$ be an auxiliary potential, such that there exists
a solution to
\begin{equation*}
\frac{d^2 \phi}{d y^2} + 4
\frac{d \phi}{d y}
- \frac{\d V_0}{\d \phi} = 0 \; ,
%\label{sep15-14-5}
\end{equation*}
that starts at $y \to -\infty$ in the true vacuum
$\phi = \phi_+$ and reaches the false vacuum $\phi = 0$ at
$y \to \infty$. A necessary condition for the
existence of such a sulution is again  $V_0^{\prime \prime}<- 4$
at the maximum of $V_0$.
Let the potential $V(\phi)$ be a deformation of $V_0$ which is
deeper than
$V_0$ in the true vacuum, see Fig.~\ref{potential}.
Then, by continuity, there exists
a value of $\phi_* <  \phi_+$ such that
the solution to  eq.~\eqref{neweqm} starts at
$s =0$ from
$\phi_c = \phi_*$
and approaches the false vacuum
$\phi =0$ as $s \to \infty$: a solution starting close to
the top of $(-V)$ overshoots the false vacuum, while a solution
that starts from small $\phi$ undershoots it. This completes the
argument.

The bounce solution approaches the false vacuum $\phi=0$ exponentially
in $s$,
\be
\phi_c\to c \e^{- \Delta_+ s} \simeq \frac{c}{(2\cosh s)^{\Delta_+}}\; , \;
\; \; \;\;\;\; c = \mbox{const} \; , \;\;\; s \to \infty \; ,
\label{jan14-15-4}
\ee
where $\Delta_+$ is given by eq.~\eqref{jan14-15-1} and
$m^2 = V^{\prime \prime} (0)$.
We find from eq.~\eqref{jan13-15-3} that
\be
\cosh s = \frac{1 + z^2 + R^2}{2z} \equiv  \frac{1 + z^2 + x^2}{2z}\; .
%\label{jan14-15-3}
\label{jan15-15-1}
\ee
Hence, $s = \infty$ corresponds to the boundary, and near the boundary
\[
\phi_c = c \left( \frac{z}{x^2 + 1} \right)^{\Delta_+} \; .
\]
We compare this with eqs.~\eqref{dec10-14-1}, \eqref{dec10-14-3} and find
the expectation value of an operator in the (Euclidean) boundary CFT:
\be
\langle {\cal O} \rangle  = 2c\sqrt{m^2 + 4} \cdot
\frac{1}{(x^2 + 1)^{\Delta_+}} \; .
\label{dec14-15-7}
\ee
This is the CFT analog of the Fubini--Lipatov bounce of unit size,
cf. eq.~\eqref{dec11-14-1}.

To construct the bounce of arbitrary size, we make yet another change of
coordinates,
\be
\e^u = \sqrt{z^2 + R^2} \; , \;\;\;\;\;\;
\cos v = \frac{z}{\sqrt{z^2 + R^2}} \; ,
\label{jan14-14-6}
\ee
so that
\be
\cosh s = \frac{\cosh u}{\cos v} \; , \;\;\;\;
\sinh s \cdot \cos \psi = \frac{\sin v}{\cos v} \; .
\label{jan14-15-10}
\ee
In terms of these coordinates, the metric is
\be
ds_5^2 = \frac{1}{\cos^2 v} \left(du^2 + dv^2 + \sin^2 v d\Omega_3^2 \right)
\; .
\label{jan14-15-2}
\ee
Let us write for the bounce of unit size
\be
\phi_c (s) = f (\cosh s) = f \left( \frac{\cosh u}{\cos v} \right) \; .
\label{jan14-15-5}
\ee
Since the metric \eqref{jan14-15-2} is invariant under translations of
$u$, there is a family of bounce solutions parametrized by a parameter $b$:
\[
\phi_c ^{(b)} =  f \left( \frac{\cosh (u + b)}{\cos v} \right) \;
\]
(we keep the notation $\phi_c$ for the bounce of unit size, i.e.,
$\phi_c \equiv \phi_c^{(b=0)}$).
The bounce   $\phi_c^{(b)}$ has size
\[
\rho = \e^{-b} \; .
\]
One way to see this is
to consider the behavior of $\phi_c^{(b)}$ near the boundary.
Making use of eqs.~\eqref{jan14-15-4} and \eqref{jan14-15-5}
we find that $f(C) = c/(2C)^{\Delta_+}$ at large $C$, and hence
near the boundary
\[
\phi_c^{(b)} = \frac{c}{\left[2  \cosh (u + b)/\cos v \right]^{\Delta_+}}
\; ,
\]
which, in view of eq.~\eqref{jan14-14-6}, gives
\be
\phi_c ^{(b)} = c z^{\Delta_+}\cdot \frac{\e^{- \Delta_+ b}}{\left(R^2
+ \e^{-2b}\right)^{\Delta_+}} =  c z^{\Delta_+} \left(\frac{\rho}{x^2 + \rho^2}
\right)^{\Delta_+} \; , \;\;\;\;\; z \to 0 \; .
\label{feb11-15-5}
\ee
This shows that $\rho$ is indeed the size of the bounce.

Another way to see that $b$ parametrizes the size of the bounce is
to consider lines of constant $\phi_c ^{(b)}$ in the $(z,R)$-plane.
Line of constant $\phi_c^{(b)} = f(C)$ obeys
\[
\frac{\cosh (u+b)}{\cos v} = C \; .
\]
For given $C$, these lines are different for different $b$, while
the value of $\phi_c^{(b)}$ is the same for all $b$.
Making use of eq.~\eqref{jan14-14-6}, we find the equation of the line
in $(z,R)$-coordinates
\[
(z- C\e^{-b})^2 + R^2 = (C^2 - 1) \e^{-2b} \; .
\]
It shows that these lines
are circles (4-spheres in coordinates $(z, x^\mu)$)
of radii $\sqrt{C^2 -1} \e^{-b}$ centered at
$z = C\e^{-b}$, cf. Ref.~\cite{Gorsky}.
As compared to the bounce of unit size ($b=0$),
the center of the 5-dimensional bounce (the point where $C=1$)
is shifted
from $z=1$ to $z=\e^{-b} = \rho$, and the bounce configuration is
dilated by a factor $\e^{-b} = \rho$. This is precisely what one expects
for the bounce of size $\rho$ in the holographic picture.

To end up the discussion of the Euclidean bounce, we derive the infinitesimal
form of its scale transformation. We write for the transformation of
the bounce of unit size
\[
\delta_b \phi \equiv \frac{\d \phi_c ^{(b)}}{\d b} (b=0) =
\frac{df(C)}{dC} ~\frac{\sinh u}{\cos v} = \frac{d \phi_c}{d s}
 \frac{\sinh u}{\sinh s \cos v} \; .
\]
Making use of eq.~\eqref{jan14-15-10} we obtain finally
\be
\delta_b \phi =  \frac{d \phi_c}{d s} \sin \psi \; .
\label{jan19-15-1}
\ee
We will use this result in what follows.

\subsection{Expanding bubble and its radial perturbations}
\label{subsec:adSradial}

Without loss of generality,
from now on we concentrate on the bounce of unit size.
To study the bounce and its perturbations in adS$_5$ with Minkowski
signature, we analytically continue the coordinate $\psi$ in
eq.~\eqref{jan13-15-2},
\[
\psi \to i \tau \; ,
\]
so the metric becomes
\begin{align}
%\label{oct29-14-2}
ds^2_5&=ds^2+\sinh^2 s(-d\tau^2+\cosh^2\tau d\Omega_3^2)
\nonumber \\
&= ds^2+\sinh^2 s [-d\tau^2+\cosh^2\tau (d\chi^2 + \sin^2 \chi~d\Omega_2^2)]\;.
\label{oct29-14-2}
\end{align}
This is the metric of adS$_5$ with dS$_4$ slicing (see, e.g.,
Ref.~\cite{Goon:2011qf}): the hypersurfaces $s = \mbox{const}$
are dS$_4$ spaces with metric proportional to
\be
ds_4^2    = -d\tau^2+\cosh^2\tau d\Omega_3^2 \equiv
g_{4\, \mu \nu} dX^\mu dX^\nu \; .
\label{oct29-14-5}
\ee
The region near the adS$_5$ boundary is conveniently
studied using the Poincar\`e coordinates, in which the metric has the
standard form
\[
ds_5^2 = \frac{1}{z^2}[-(dx^0)^2 + dz^2 +
d{\cal R}^2 + {\cal R}^2 d\Omega_2^2] \; ,
\]
where ${\cal R}$ is the usual radial coordinate in 3d space.
The transformation to these coordinates is
\begin{subequations}
\begin{align}
z&= \frac{1}{\cosh s + \sinh s \cosh \tau \cos \chi}
\nonumber \\
x^0 &=  \frac{\sinh s \sinh \tau}{\cosh s + \sinh s \cosh \tau \cos \chi}
\nonumber \\
{\cal R} &
=  \frac{\sinh s \cosh \tau \sin \chi }{\cosh s + \sinh s \cosh \tau \cos \chi}
\; ,
\nonumber
\end{align}
\end{subequations}
or
\begin{subequations}
\begin{align}
\cosh s = \frac{{\cal R}^2 - (x^0)^2 + z^2 +1}{2z}
&\equiv \frac{x^2 + z^2 +1}{2z}
\label{oct30-14-2}
\\
\sinh s \sinh \tau = \frac{x^0}{z} \; ,
\;\;\;\; \;\;\;\,\;\;\;\; &
\label{oct30-14-5}
\\
\sinh s \cosh \tau \sin \chi = \frac{\cal R}{z} \; ,
 \label{oct30-14-6}
\end{align}
\end{subequations}
where $x^2 = \eta_{\mu \nu}x^\mu x^\nu$.
The hypersurface $s=0$ is the light cone  emanating from $z=1$,
$x^\mu = 0$, so
the coordinates used here cover the exterior of this light cone only;
since we will be eventually interested in the behavior of the fields
near the
adS$_5$ boundary, we will not need to consider the interior of this
light cone.

The bounce solution $\phi_c (s)$ is the same function
of $s$ as before; it obeys eq.~\eqref{neweqm}. Since eq.~\eqref{oct30-14-2}
formally coincides with eq.~\eqref{jan15-15-1}, the expectation value of the
CFT operator is still given by eq.~\eqref{dec14-15-7}, now with Minkowski
$x^2$. It diverges as $x^2 \to -1$; in the holographic picture this is
the surface at which the light cone $s=0$ hits the adS$_5$ boundary.
According to the general scenario of the (pseudo-)conformal Universe,
we assume that the rolling stage terminates before that.

Let us now consider the perturbations about the expanding bubble,
$\phi = \phi_c (s) + \delta \phi (s, X^\mu)$,
where $X^\mu$ are coordinates on dS$_4$, see eq.~\eqref{oct29-14-5}.
We still call $\delta \phi$ radial perturbations.
In metric \eqref{oct29-14-2}
the quadratic action is
\[
S_2= \int~ds~d^4X~\sinh^4s \sqrt{-g_4} \left[-\frac{1}{2}
\left(\d_s \delta \phi\right)^2
- \frac{1}{2 \sinh^2 s} g^{\mu \nu}_4
\d_\mu \delta \phi \cdot \d_\nu \delta \phi - \frac{1}{2}
V^{\prime \prime} (\phi_c) \cdot (\delta \phi)^2 \right] \; ,
\]
and the field equation reads
\be
\d_s^2 \delta \phi + 4 \frac{\cosh s}{\sinh s} \delta \phi
+ \frac{1}{\sinh^2 s} \Box_4 \delta \phi - V^{\prime \prime} (\phi_c)
\delta \phi = 0 \; ,
\label{jan15-15-2}
\ee
where $\Box_4$ is the d'Alembertian in the
4d de~Sitter space with metric \eqref{oct29-14-5}.
Solutions to eq.~\eqref{jan15-15-2} are linear combinations of
\be
\delta \phi_{(\mu)} = \Phi_{(\mu)} (s) \Psi_{(\mu)} (X) \; ,
\label{feb11-15-3}
\ee
where
\be
\Box_4 \Psi_{(\mu)} = \mu^2 \Psi_{(\mu)} \; ,
\label{oct29-14-3}
\ee
and $\Phi_{(\mu)} (s)$ obeys the following equation:
\be
\frac{d^2 \Phi_{(\mu)}}{ds^2} + 4 \frac{\cosh s}{\sinh s} \frac{d
\Phi_{(\mu)}}{ds} + \frac{\mu^2}{\sinh^2 s} \Phi_{(\mu)} - V^{\prime \prime}
(\phi_c) \Phi_{(\mu)} = 0 \;.
\label{oct30-14-1}
\ee
We will get back to eq.~\eqref{oct30-14-1} shortly, and now we
discuss solutions to eq.~\eqref{oct29-14-3}.
The latter is the equation for a field of mass $\mu$
in 4d de~Sitter space-time \eqref{oct29-14-5}.
We are again interested in
short modes, i.e., modes whose conformal 3d momentum is high, $k \gg 1$.
The interesting regime is then $\tau \gg 1$, and the spatial curvature
of slices of constant $\tau$ can be neglected. Working near $\chi=0$
without loss of generality, we approximate the metric \eqref{oct29-14-5}
by the metric of
dS$_4$ with flat slices of constant time,
\[
ds_4^2 = - d\tau^2 + \left(\frac{\e^{\tau}}{2}\right)^2
d{\bf x}^2 = \frac{1}{\eta^2}
(-d\eta^2 + d{\bf x}^2) \; ,
\]
where
\be
\eta = -2 \e^{-\tau} \; .
\label{jan18-15-4}
\ee
Upon introducing the field
$\sigma_{(\mu)} = \Psi_{(\mu)}/\eta$, the field equation \eqref{oct29-14-3}
becomes
\[
- \d_\eta^2 \sigma_{(\mu)} + \Delta \sigma_{(\mu)} + \frac{2-\mu^2}{\eta^2}
\sigma_{(\mu)} = 0 \; ,
\]
where $\Delta$ is the flat 3d Laplacian. In 3d momentum representation
its solutions tending to $e^{-ik\eta}/\sqrt{2k}$ as $\eta \to -\infty$
are
\[
\sigma_{(\mu)} = \frac{\sqrt{\pi |\eta|}}{2}  H_\nu^{(1)} (-k\eta) \; ,
\]
where
\[
\nu = \sqrt{\frac{9}{4} - \mu^2} \; .
\]
For real $\nu$
the asymptotics of these solutions as $k |\eta| \to 0$ are
\[
\sigma_{(\mu)} = -i \frac{\sqrt{\pi} 2^{\nu-1}}{\Gamma(1-\nu) \sin{\nu \pi}}
\frac{1}{k^\nu (-\eta)^{\nu-1/2}}\; ,
\]
and the  late-time asymptotics of $\Psi_{(\mu)}$ are
\be
\Psi_{(\mu)} = -i \frac{\sqrt{\pi} 2^{\nu-1}}{\Gamma(1-\nu) \sin{\nu \pi}}
\frac{1}{k^\nu (-\eta)^{\nu-3/2}}\; , \;\;\;\;\;\;\;\; k|\eta| \ll 1 \; .
\label{jan18-15-2}
\ee
For $\nu > 1/2$ the modes $\sigma_{(\mu)}$ are
enhanced at $k|\eta| \ll 1$ as compared to massless
Minkowski theory, in which $g \propto k^{-1/2}$. In what follows
we consider the case
\[
\mu^2<2 \; , \;\;\;\;\;\;\; \nu > 1/2 \; .
\]
In this case the quantum field associated with the mode of mass $\mu$
can be considered as classical random field~\cite{Polarski:1995jg}
with power spectrum
${\cal P}_{(\mu)} \propto k^{-2\nu + 3}$.

Let us now turn to eq.~\eqref{oct30-14-1}. For $\mu^2 < 2$
this is an eigenvalue
equation for $\mu^2$.
We normalize its solutions as follows:
\be
\int~ds~\sinh^2s~|\Phi_{(\mu)}|^2 =1 \; ,
\label{oct30-14-3}
\ee
then the field $\Psi_{(\mu)} (X)$ has canonical kinetic term.
The eigenfunctions must behave at large $s$ (near the adS$_5$ boundary)
as follows,
\be
\Phi_{(\mu)} = C_{(\mu)} \e^{-\Delta_+ s}
\label{jan18-15-1}
\ee
(another
asymptotic behavior $\Phi_{(\mu)} \propto \e^{-\Delta_- s}$
with $\Delta_- = 2 - \sqrt{4+m^2}$
would correspond to a non-zero source at the boundary),
where $C_{(\mu)}$ is determined by the normalization condition
\eqref{oct30-14-3}.
Near $s=0$ the solutions are $\Phi_{(\mu)} \propto s^{n_\pm}$, where
$n_{\pm} = - \frac{3}{2} \pm \sqrt{\frac{9}{4} - \mu^2}$.
For $\mu^2 < 2$ one of them is normalizable, and another is
not. Hence, we are indeed dealing with an eigenvalue problem.

Let us consider the behavior of the modes near the boundary $s\to \infty$
and at
late times.
We are interested in short modes and, as before,
work in the vicinity of $\chi = 0$.
Making use of
eq.~\eqref{oct30-14-2} we translate eq.~\eqref{jan18-15-1} into
\[
\Phi_{(\mu)} = C_{(\mu)} \frac{z^{\Delta_+}}{(x^2 + 1)^{\Delta_+}} \; ,
\;\;\;\;\;\; z \to 0 \; .
\]
We recall eqs.~\eqref{jan18-15-4}
and \eqref{oct30-14-5} to write at large $s$
\[
-\eta = \frac{x^2 + 1}{2 x^0} \; .
\]
Furthermore, our region of small $\chi$ and $x^2 \to -1$ corresponds to
$x^0 \to 1$, ${\cal R} \ll x^0$, see eqs.~\eqref{oct30-14-5}
and \eqref{oct30-14-6}. We use eq.~\eqref{jan18-15-2} and get finally
\[
\delta \phi_{(\mu)} = -i C_{(\mu)}
\frac{\sqrt{\pi} 2^{2\nu- 5/2}}{\Gamma(1-\nu) \sin{\nu \pi}}
\cdot \frac{z^{\Delta_+}}{(x^2 + 1)^{\Delta_+ + \nu - 3/2}} \cdot
\frac{1}{k^\nu} \cdot B_{\bf k}  + \mbox{h.~c.}\; , \;\;\;
\;\;\; z \to 0 \; , ~k(x^2 + 1) \ll 1 \; ,
\]
where $B_{\bf k}$ is annihilation operator.
The radial perturbations
in the boundary CFT are obtained from
eqs.~\eqref{jan14-15-1},
\eqref{dec10-14-3}:
\be
\delta {\cal O}_{(\mu)}  = -i C_{(\mu)} \sqrt{m^2 + 4}
\frac{\sqrt{\pi} 2^{2\nu- 3/2}}{\Gamma(1-\nu) \sin{\nu \pi}}
\cdot \frac{1}{(x^2 + 1)^{\Delta_+ + \nu - 3/2} \cdot
k^\nu} \cdot B_{\bf k}  + \mbox{h.~c.}\; ,
\;\; z \to 0 \; ,  ~k(x^2 + 1) \ll 1 \; .
\label{feb11-15-1}
\ee
For $\nu > 1/2$ these modes have enhanced power spectrum,
so they may lead to interesting
effects beyond the linear level.

Now, the key  point out that eq.~\eqref{oct30-14-1} always admits a
solution with
\[
\mu^2 = -4 \; .
\]
This solution is
\be
\Phi_{(\mu^2 = -4)} = \frac{d \phi_c}{ds} \; .
\label{feb11-15-2}
\ee
One can view its existence as a consequence of the dilatational
invariance. Indeed, this invariance guarantees that the function
\eqref{jan19-15-1} is a solution to the equation for perturbations
in the Euclidean domain, its counterpart in Minkowski signature
being
\[
\delta_b \phi = \frac{d \phi_c}{ds} \sinh \tau \; .
\]
This solution has the form \eqref{feb11-15-3}, and the 4d factor
$\sinh \tau$ obeys eq.~\eqref{oct29-14-3} with $\mu^2 = -4$. Hence,
 $d \phi_c/ds$ must obey eq.~\eqref{oct30-14-1} with $\mu^2=-4$, and it
indeed does. Note that the solution  \eqref{feb11-15-2} does not have nodes,
which implies that $\mu^2 = -4$ is the lowest eigenvalue of
eq.~\eqref{oct30-14-1}. Therefore, the leading asymptotics  of
$\delta {\cal O}$ at late times ($x^2 \to -1$)
is determined by the mode with $\mu^2 = -4$.

For $\mu^2 = -4$ we have $\nu = 5/2$, and the late-time asymptotics
of the CFT operator \eqref{feb11-15-1} reads
\be
\delta {\cal O}_{(\mu^2=-4)}  = -i C_{(\mu^2=-4)}  \sqrt{m^2 + 4} \cdot 3 \cdot 2^{3/2}
\cdot \frac{1}{(x^2 + 1)^{\Delta_+ + 1} \cdot
k^{5/2}} \cdot B_{\bf k}  + \mbox{h.~c.}\; .
\label{feb11-15-4}
\ee
In complete analogy to eq.~\eqref{dec17-14-1},
these radial perturbations in the boundary CFT have red power spectrum
${\cal P}_{\delta {\cal O}} \propto k^{-2}$.
In view of eq.~\eqref{feb11-15-5} and by the same argument as in
Section~\ref{subsec:radial},
the dependence on $x$ in \eqref{feb11-15-4} can be interpreted
as the effect of spatially varying scale $\rho({\bf x})$. This
is the desired result.

\subsection{Dimension-zero field: phase}
\label{subsec:adSphase}

Let us extend the model by promoting the field $\phi$ to a complex
scalar field. With the kinetic term $(1/2)\d \phi^* \d \phi$ and
$V= V(|\phi|)$, the model has now a global $U(1)$ symmetry.
We take the bounce/bubble solution $\phi_c$
real, then away from $\phi=0$,
the phase $\mbox{Arg}~\phi$ corresponds to a dimension-zero
operator, the phase of ${\cal O}$. To study its late-time
perturbations, we consider the field  $\varphi = \mbox{Im} ~\phi$.
The quadratic action is
\[
S_2= \int~ds~d^4X~\sinh^4s \sqrt{-g_4} \left[-\frac{1}{2} \left(\d_s
\varphi \right)^2 - \frac{1}{2 \sinh^2 s} g^{\mu \nu}_4 \d_\mu \varphi
\d_\nu \varphi - \frac{1}{2} \frac{V^{\prime} (\phi_c)}{\phi_c} \varphi^2
\right] \; .
\]
The solutions to the field equation again have the product form
\eqref{feb11-15-3} where the 4d part obeys eq.~\eqref{oct29-14-3}, while
the equation for the $s$-dependent factor, which we denote by
$\Phi_{I,(\mu)}$, is
\be
\d_s^2 \Phi_{I, (\mu)} + 4 \frac{\cosh s}{\sinh s} \Phi_{I, (\mu)}
+ \frac{\mu^2}{\sinh^2 s} \Phi_{I, (\mu)} - \frac{V^{\prime}
(\phi_c)}{\phi_c} \Phi_{I,(\mu)} = 0 \; .
\label{oct30-14-10}
\ee
It
has a solution with $\mu =0$; this is $\Phi_{I,(\mu=0)} = \phi_c$. The
existence of this mode is guaranteed by the $U(1)$-symmetry: both $\phi_c$
and $\e^{i\alpha} \phi_c$ are solutions to the full non-linear field
equation, so $i\alpha \phi_c$ must be a solution for the linearized
equation for perturbations for $k=0$. The solution  $\Phi_{I,(\mu=0)} =
 \phi_c$ again does not have nodes, so it is the lowest eigenmode of
eq.~\eqref{oct30-14-10} which determines the leading late-time asymptotics
of $\mbox{Im}~{\cal O}$.

For $\mu^2=0$ we have $\nu = 3/2$, and eq.~\eqref{feb11-15-1} gives
\[
\mbox{Im}~{\cal O} = -i   C_{(\mu=0)} \sqrt{2(m^2 + 4)}
\cdot \frac{1}{(x^2 + 1)^{\Delta_+} \cdot
k^{3/2}} \cdot A_{\bf k}  + \mbox{h.~c.}\; ,
\;\; z \to 0 \; ,  ~k(x^2 + 1) \ll 1 \; ,
\]
where $A_{\bf k}$ is another annihilation operator.
Thus, the phase $\Theta = \mbox{Im}~{\cal O}/\langle {\cal O}\rangle$
freezes out at late times at
\[
\Theta = -i \frac{C_{(\mu=0)}}{\sqrt{2}c}\cdot \frac{1}{k^{3/2}}
\cdot A_{\bf k}  + \mbox{h.~c.} \; ,
\]
where the constant $c$ is determined by the bounce solution via
eq.~\eqref{jan14-15-4}. In complete analogy to Section~\ref{dim0},
the phase perturbations have flat power spectrum
\[
{\cal P}_\Theta = \frac{1}{4\pi^2} \left( \frac{C_{(\mu=0)}}{c} \right)^2
\; .
\]
We conclude that our holographic model shares all late-time properties
of the 4d \\(pseudo-)conformal models.

\section{Conclusion}
\label{sec:concl}

While the overall picture of the (pseudo-)conformal Universe
is the same in the false vacuum decay and original homogeneous rolling models,
the former may have peculiarities. First, the slices on which
the background is homogeneous, are not spatially flat. Depending on
the embedding of the (pseudo-)conformal mechanism into a complete
cosmological scenario, this may or may not give rise to novel
properties of the resulting adiabatic perturbations, such
as particular form of
statistical anisotropy. Second, in the holographic model
there may exist modes, which are subdominant at late times
(with $\mu^2 > -4$ and $\mu^2 >0$ for radial and phase perturbations,
respectively), but nevertheless relevant. Also, back reaction of
the bounce and its perturbations on space-time metric, which we
neglected in this paper, may possibly lead to interesting effects.
We leave the analysis of these and other issues for future work.

\vspace{0.3cm}

We are indebted to R.~Gopakumar, R.~Rattazzi, S.~Sibiryakov and A.~Starinets
for useful discussions.
The work of M.L. and V.R. has been supported by Russian Science Foundation
grant  14-22-00161.

\end{document}